\documentclass[12pt]{iopart}
\usepackage{graphicx}
\usepackage{subfigure}
\usepackage{amssymb}

\newcommand {\snn}      {\sqrt{s_{_{\rm NN}}}}

\newcommand {\pT}		{p_{\rm T}}

\newcommand {\deta}     {\Delta\eta}
\newcommand {\dphi}     {\Delta\phi}

\begin{document}
\title[Heavy-flavor correlations in STAR]{Heavy-flavor correlation measurements via 
electron azimuthal correlations with open charm mesons}
\author{Andr\'e Mischke for the STAR Collaboration \footnote{For the full author list, see \cite{Star:npe}.}}
\address{Institute for Subatomic Physics, Utrecht University, Princetonplein 5, 3584 CC Utrecht, the Netherlands.}
\ead{a.mischke@phys.uu.nl}
\begin{abstract}
We report first STAR measurement on two heavy-flavor particle correlations in p+p collisions at RHIC. 
Heavy-flavor (charm and bottom) events are identified and separated on a statistical basis by the azimuthal correlation of their decay electrons and open charm mesons, which yield important information about the underlying production mechanism. 
The azimuthal correlation distribution exhibits a two-peak structure which can be attributed to $B$ decays on the near-side and predominantly charm pair production on the away-side. These assumptions are supported by dedicated simulations using PYTHIA and MC@NLO event generators. This novel correlation technique will allow comprehensive energy-loss measurements of heavy quarks in heavy-ion collisions.
\end{abstract}


\section{Introduction}
The investigation of heavy-flavor production in heavy-ion collisions provides key tests of parton energy-loss models and, thus, yields important information about the properties of the produced highly-dense QCD medium. Due to their large mass ($m >$ 1.2 GeV/$c^2$), heavy quarks are produced primarily in the early stage of the collision by hard scattering processes (large momentum transfer) and probe the complete space-time evolution of the medium. As recent RHIC measurements have shown~\cite{Star:yfei}, heavy-quark production by initial state gluon fusion also dominates in heavy-ion collisions where many, in part overlapping nucleon-nucleon collisions occur. Heavy-quark production by thermal processes later in the collision is low since the expected energy available for particle production in the medium ($\approx$0.5 GeV) is smaller than the energy needed to produce a heavy-quark pair ($>$2.4 GeV). Interaction processes of heavy quarks can be calculated in pQCD~\cite{Frix:pqcd} and their yields are sensitive to the initial gluon density~\cite{hq:prod}. Theoretical models based on perturbative QCD predicted~\cite{El:deadcone,El:magd} that heavy quarks should experience smaller energy loss in the medium than light quarks when propagating through the extremely dense medium due to the mass-dependent suppression (called "dead-cone effect").
Charm and bottom mesons are currently identified by assuming that isolated electrons in the event stem from semi-leptonic decays of heavy-quark mesons. At large transverse momentum ($\pT$), this mechanism of electron production is dominant enough to reliably subtract other sources of electrons (conversions from photons and Dalitz decays). STAR measurements in central Au+Au collisions have shown~\cite{Star:npe} that the electron yield from semi-leptonic heavy-quark decays exhibits an unexpectedly large suppression, suggesting substantial energy loss of heavy quarks in the produced medium. Surprisingly, the amount of suppression at high $\pT$ is at the same level as observed for light-quark hadrons, which was not expected due to the dead-cone effect. Energy-loss models incorporating contributions from charm and bottom do not describe the observed suppression sufficiently (for a detailed discussion, see~\cite{Star:npe}). Although it has been realized that energy loss by parton scattering is probably of comparable importance to energy loss by gluon radiation, the quantitative description of the data is still not satisfying. The remaining discrepancy between data and model calculations could indicate that the $B$ dominance over $D$ mesons starts at a higher $\pT$ as expected. To verify this assumption an urgent need arises to disentangle the $D$ and $B$ contribution to the non-photonic electron distribution experimentally. In this paper, we present a novel analysis technique to identify and separate charm and bottom quark events via leading electron azimuthal correlations with open charm mesons. The specific advantage of this correlation method, in contrast to the conventional heavy-quark measurements, is the possibility to efficiently trigger on heavy-quarks using their decay electrons. The STAR electromagnetic calorimeter provides a unique opportunity to identify electrons on the trigger level and, therefore, to select a sample of events with a large enhancement of heavy-flavor production. Moreover, this correlation method reduces significantly the combinatorial background in the reconstruction of $D^0$ mesons.

\section{Heavy-flavor correlations}
\begin{figure}[t]
\begin{center}
\subfigure{\includegraphics[width=0.4\textwidth]{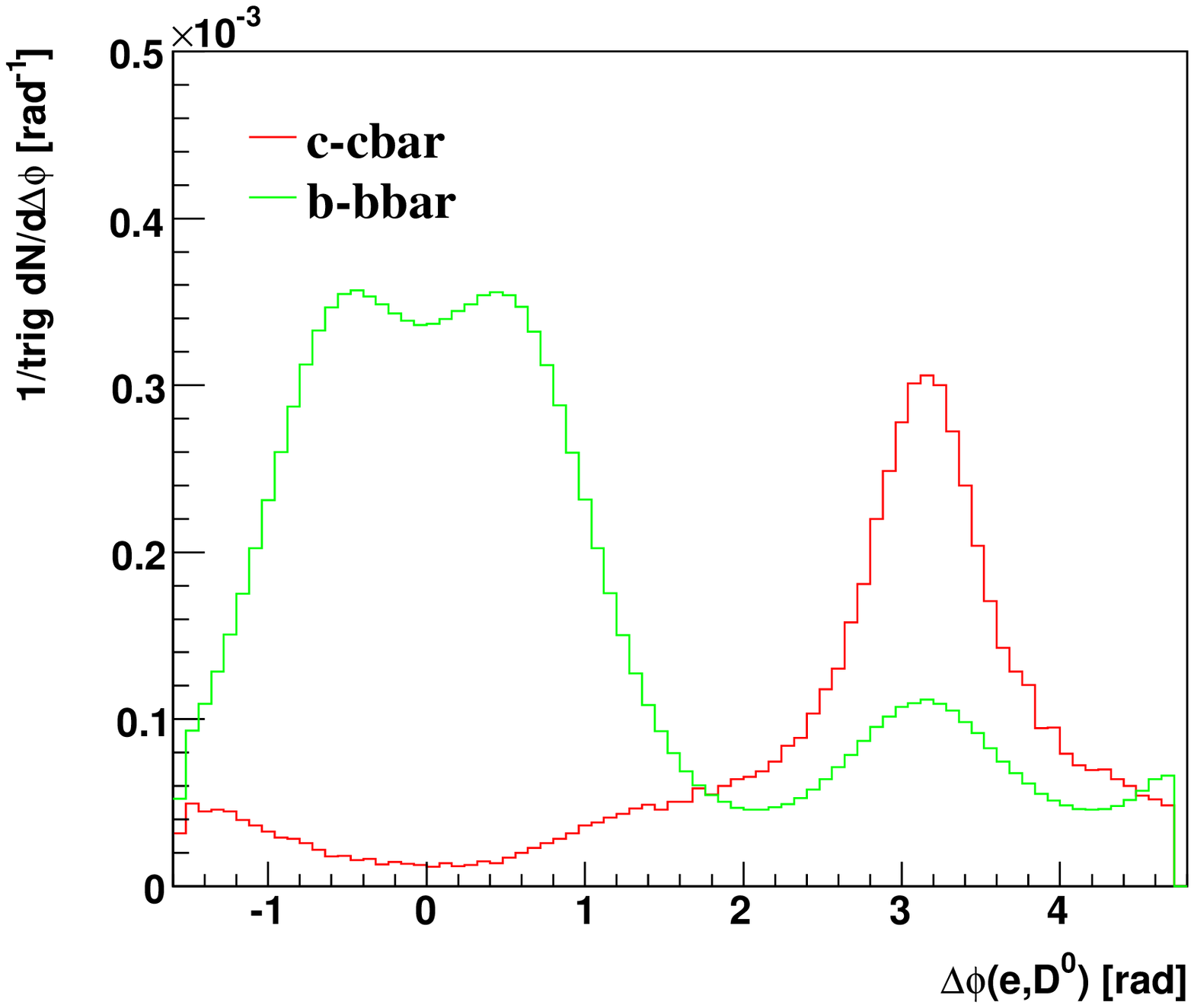}}
\hspace*{1.cm}
\subfigure{\includegraphics[width=0.4\textwidth]{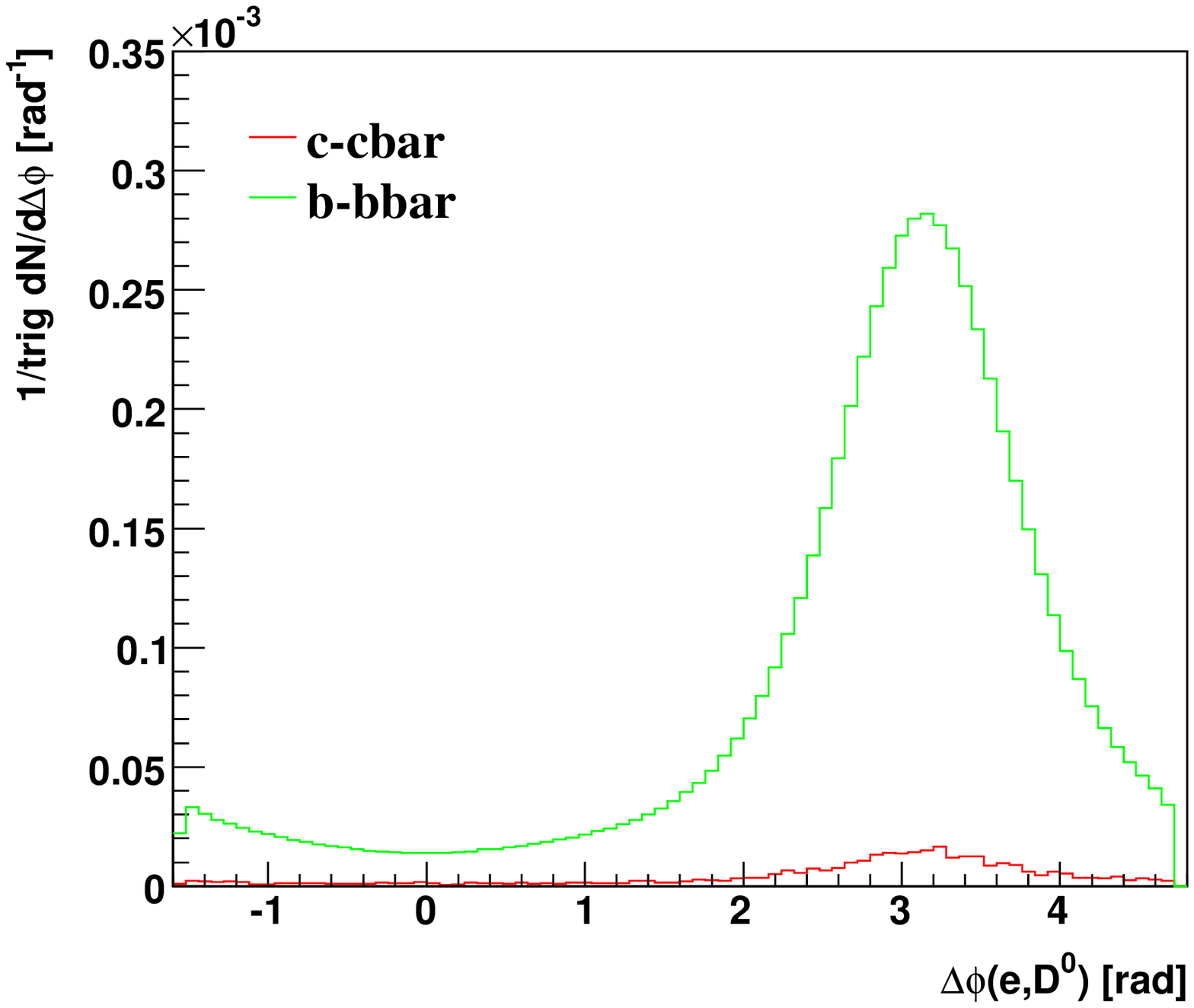}}
\vspace*{-0.5cm}
\caption[]{\label{fig:1}
PYTHIA simulations for the azimuthal angular correlation distribution of non-photonic electrons and $D^0$ mesons for like-sign (left panel) and unlike-sign (right panel) electron$-$Kaon pairs. The red (green) histogram illustrates the charm (bottom) contribution for trigger electrons with $\pT > 3$ GeV/c.}
\end{center}
\end{figure}

Flavor conservation implies that heavy quarks are always produced in quark anti-quark pairs. A more detailed understanding of the underlying production process can be obtained from events in which both heavy-quark particles are detected. Due to momentum conservation, these heavy-quark pairs are correlated in relative azimuth ($\dphi$) in the plane perpendicular to the colliding beams, leading to the characteristic back-to-back orientated sprays of particles.
This correlation survives the fragmentation process to a large extent in p+p collisions. 
In this analysis, charm and bottom production events are identified using the characteristic decay topology of their jets. 
Charm quarks predominantly ($\approx$54$\%$) hadronize directly and bottom quarks via $B$ decays into $D^0$ mesons. The branching fraction for charm and bottom decays into electrons is $\approx$10$\%$. While triggering on the so-called leading electron (trigger side), the balancing heavy quark, which is identified by the $D^0$ meson, can be used to identify the underlying production mechanisms (probe side). 
In addition, a charge-sign requirement on the trigger electron and decay Kaon provides a powerful tool to separate charm and bottom quark events. Fig.~\ref{fig:1} illustrates the azimuthal correlation distribution for like- (left) and unlike-sign electron$-$Kaon pairs (right) obtained from PYTHIA simulations. On the left panel of Fig.~\ref{fig:1}, the near-side peak is dominated by $D^0$ mesons from $B$ decays whereas the away-side peak stems mainly from charm pair production (flavor creation). By contrast, the away-side peak for unlike-sign electron$-$Kaon pairs (right panel) originates essentially from $B$ decays only. Thus, the azimuthal correlation in combination with the charge-sign requirement allows the clear separation of charm and bottom quark events.

\section{Data analysis}
The analysis is based on Run VI p+p data taken at $\snn$~=~200 GeV by the STAR experiment at RHIC. Run VI provided the first dataset where the calorimeter was fully installed and operational. The integrated luminosity was 9 pb$^{-1}$, of those 1.2 million events were used after a cut on the z-coordinate of the collision vertex (beam axis) within 30~cm of the TPC center. The tight vertex cut is used to minimize the amount of material within the detector volume causing photon conversions.

Particle identification via ionization energy loss and tracking over a large kinematical range with very good momentum resolution is performed by the Time Projection Chamber (TPC)\cite{Star:tpc} . The TPC 
has an acceptance of $|\eta| < 1.4$ and full azimuthal coverage. The STAR detector utilizes a barrel-electromagnetic calorimeter (BEMC)\cite{Star:emcal} as a leading particle (electrons and photons) trigger to study high $\pT$ particle production. 
The calorimeter, situated behind the TPC, covers an acceptance of $|\eta| < 1$ and full azimuth. To enhance the high $\pT$ range, a high-tower trigger was used with an energy threshold of 5.4~GeV for the highest energy in a BEMC cell. The high-tower trigger efficiency is nearly 100$\%$. 

The electron identification is performed by combining the information from the TPC and the BEMC (cell energy). Due to the finite momentum resolution of the calorimeter, only particles with $\pT > 1.5$ GeV/c can be measured. 
A shower maximum detector, located at a depth of 5 radiation length inside the calorimeter modules, measures the profile of an electromagnetic shower and the position of the shower maximum with high resolution ($\deta, \dphi$) = (0.007, 0.007). In contrast to hadrons, electrons deposit most of their energy in the BEMC cells. A cut on the shower profile size combined with a requirement on the ratio momentum-to-cell energy, $0 < p/E < 2$, reject a large amount of hadrons. The final electron sample is obtained by applying a momentum dependent cut on the ionization energy loss ($3.5 < dE/dx < 5.0$~keV/cm). The resulting hadron suppression factor is 10$^5$ at $\pT$ = 2 GeV/c and 10$^2$ at $\pT$ = 7 GeV/c. The electron purity is $\approx$100$\%$ up to $\pT$ = 5 GeV/c and decreases to 97$\%$ at $\pT$ = 7 GeV/c.
Most of the electrons in the final state are originating from other sources than heavy-flavor decays. Photon conversions ($\gamma \rightarrow e^+e^-$) in the detector material between the interaction point and the TPC and neutral pion and $\eta$ Dalitz decays ($\pi^0, \eta \rightarrow e^+e^- \gamma$) represent the dominant source of the so-called photonic electrons. Contributions from other decays, like $\rho$, $\phi$ and Ke3, are small and can be neglected. In this analysis, photonic electrons are identified and rejected based on invariant mass. Here, each electron candidate is combined with tracks which pass loose cuts on the ionization energy loss to preselect electron candidates~\cite{Star:npe}. Electrons with an invariant mass of $m < 150$~MeV/c$^2$ are disregarded. The photonic background finding efficiency is estimated to be $\approx$70$\%$. The ratio inclusive-to-photonic electrons is 1.35 at $\pT$ = 3 GeV/c and increases to 1.6 at $\pT$ = 7 GeV/c. About 6k non-photonic electrons, originating mostly from heavy-flavor decays, are used for the further analysis. 
\begin{figure}[t]
\begin{center}
\includegraphics[width=0.55\textwidth]{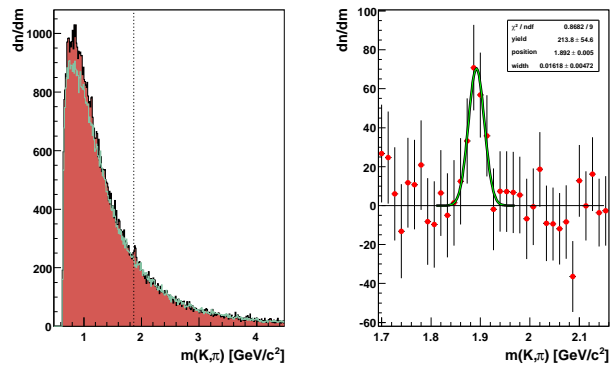}
\vspace*{-.5cm}
\caption[]{\label{fig:2}
Left panel: Invariant mass distribution of Kaon-pion pairs (red histogram) requiring a non-photonic electron trigger in the event and the combinatorial background (green histogram). The dotted horizontal line indicates the expected $D^0$ mass. Right panel: Background subtracted invariant mass distribution. The solid line is a Gaussian fit to the data around the peak region.}
\end{center}
\end{figure}
The associated $D^0$ mesons are reconstructed via their hadronic decay channel $D^0 \rightarrow K^- \pi^+$ (B.R. 3.84$\%$) by calculating the invariant mass of all oppositely charged TPC tracks in the same event. Up to now, the precise $D^0$ decay topology can not be resolved due to insufficient tracking resolution close to the collision vertex. Moreover, negative tracks have to fullfil a $dE/dx$ cut of $\pm 3\sigma$ around the Kaon band to enhance the Kaon candidate probability. Due to the high abundance of pions in the collisions, one usually has to handle a large combinatorial background of random pairs~\cite{Star:d0dAu}. In this analysis, however, only events with a non-photonic electron trigger are used for the $D^0$ reconstruction, which suppresses the combinatorial background. Furthermore, the Kaon candidates have to have the same charge sign as the non-photonic electrons (called like-sign electron$-$Kaon pairs). 
\begin{figure}[t]
\begin{center}
\includegraphics[width=0.45\textwidth]{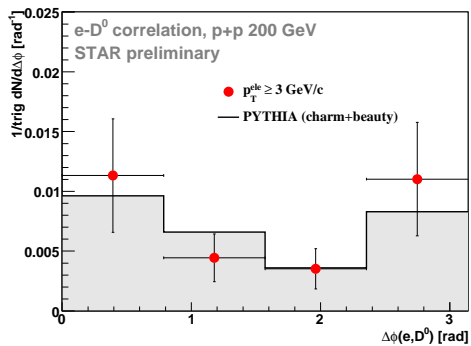}
\vspace*{-0.5cm}
\caption[]{\label{fig:3}
Azimuthal correlation distribution of non-photonic electrons and $D^0$ mesons (for like-sign electron$-$Kaon pairs) in p+p reactions at $\snn$~=~200 GeV. Statistical errors are shown only. The grey histogram illustrates results from PYTHIA simulations, which are scaled by a factor of 2.86 to match the correlation distribution for unlike-sign electron$-$Kaon pairs.}
\end{center}
\end{figure}
The resulting invariant mass distribution of Kaon-pion pairs shows a pronounced $D^0$ peak around the expected value (Fig.~\ref{fig:2}, left panel). The combinatorial background of random pairs is evaluated by combining all like-sign charged tracks in the same event. The discrepancy of the shape between the invariant mass and the combinatorial background distribution at lower masses can be understood in terms of jet particle correlations, which are not included in the background evaluation yet. It should be noted that the invariant mass distribution without a non-photonic electron trigger does not have a $D^0$ signal for the applied track quality cuts. Thus, the requirement of a non-photonic electron trigger allows suppressing the combinatorial background significantly (by a factor of $\approx$100 compared to earlier results~\cite{Star:d0dAu}), yielding a signal-to-background ratio of about 1/7 and a signal significance of about 4. 
The right panel of Fig.~\ref{fig:2} illustrates the background subtracted invariant mass distribution. The peak position and width are determined using a Gaussian fit to the data. The measured peak position, $m = 1.892 \pm$0.005 GeV/c$^2$, is slightly higher than the PDG value of 1.864 GeV/c$^2$, which can be explained by the finite momentum resolution of the TPC. The width of the signal, $\sigma_m = 16 \pm$5 MeV/c$^2$, is found to be similar to earlier results and expectations from Monte-Carlo simulations~\cite{Star:d0dAu}. Within statistical uncertainties, the $D^0$ and $\overline{D^0}$ yields are equal.

\section{Results and discussion}
The azimuthal angular ($\dphi$) correlation is calculated between the transverse momentum of the non-photonic electrons and the associated charged hadron-pairs. The Kaon-pion invariant mass distribution is obtained for different $\dphi$ bins, and the yield of the associated $D^0$ mesons is extracted as the area underneath a Gaussian fit to the signal. Fig.~\ref{fig:3} shows the azimuthal correlation distribution of non-photonic electrons and $D^0$ mesons, which exhibits a near- and away-side correlation peak with similar yields.
Comparisons to dedicated PYTHIA simulations have shown (Fig.~\ref{fig:1}, left panel) that the observed away-side correlation peak can be attributed to prompt charm pair production ($\approx$70$\%$) and $B$ decays ($\approx$30$\%$). The near-side peak, by contrast, represents essentially contributions from $B$ decays only. 
The CDF collaboration obtained a similar correlation pattern for $D^0-D^{*-}$ pairs in p+$\overline{\rm p}$ collisions at $\snn$~=~1.96 TeV~\cite{Cdf:reisert}. 
It has been shown~\cite{Field} that higher order sub-processes like gluon splitting may have a significant contribution to the near-side azimuthal correlation. 
This contribution was estimated using MC@NLO simulation~\cite{Mod:frix1}, which is a dedicated event generator with a realistic parton shower model. First results have demonstrated that the contribution from gluon splitting is small in the studied $\pT$ range.

\section{Summary and outlook}
We present first two heavy-flavor particle correlation measurement at RHIC via non-photonic electron azimuthal correlations with open charm mesons in p+p collisions at $\snn$~=~200 GeV, which allows the separation of charm and bottom production events. 
This correlation technique in combination with the STAR inner tracker system has the potential for detailed energy-loss measurement of heavy quarks in heavy-ion collisions in the future.

\ack
We are grateful to Stefano Frixione (CERN) for providing the MC@NLO code. 
This work is supported by a grant from the Netherlands Organization for Scientific Research.


\end{document}